\documentclass[prb]{revtex4}
\usepackage{graphicx}

\begin{document}
\title{Picosecond dynamics of hot carriers and phonons and scintillator non-proportionality}
\author{ A. Kozorezov, J.K.Wigmore}
\address{Department of Physics, Lancaster University,
Lancaster, LA1 4YB, UK}

\author{A.Owens}
\address{	Office for Support to New Member States, ESA/ESTEC, Noordwijk, The Netherlands}

\date{\today}

\begin{abstract}
We have developed a model describing the non-proportional response in scintillators based on non-thermalised carrier and phonon transport. We show that the thermalization of e-h distributions produced in scintillators immediately after photon absorption may take longer than the period over which the non-proportional signal forms. The carrier and LO-phonon distributions during this period remain non-degenerate at quasi-equilibrium temperatures far exceeding room temperature. We solve balance equations describing the energy exchange in a hot bipolar plasma of electrons/holes and phonons.  Taking into account dynamic screening we calculate the  ambipolar diffusion coefficient at all temperatures. The non-proportional light yields calculated for NaI are shown to be consistent with experimental data. We discuss the implications of a non-equilibrium model, comparing its predictions with a model based on the transport of thermalised carriers. Finally, evidence for non-equilibrium effects is suggested by the shape of non-proportionality curve and wide dispersion in data observed in K-dip spectroscopy near the threshold. A comparison of the predicted curves shows good agreement for deformation potential value in the range 7-8 eV.
\end{abstract}
\maketitle\onecolumngrid

\section {Introduction}
\quad The non-proportional response of light output versus energy deposition in scintillators presents a serious problem preventing otherwise good gamma-ray spectrometers from achieving a spectral resolution limited only by statistics. During the last decade considerable efforts were made in both experiment and theoretical modelling, resulting in a view that the non-proportional response originates from the non-linear interactions of electrons and holes in a tiny excitation  volume leading to a quenching of luminescence\cite{AVasiliev}. The non-proportional part of the light yield is believed to occur during a very short ($\sim$ picosecond scale) period which we will refer to below as the non-proportionality stage.

\quad The diffusion of electrons and holes during the first few picoseconds after their generation is believed to play a decisive role in the formation of a non-proportional response in scintillators. In all previous works referenced in\cite{Li} it is assumed that hot carriers thermalise on a time scale of less than 1 ps and this assumption is central to all models based on the concept  of thermalised carrier transport. While it may be valid for a few scintillating crystals  it cannot be taken for granted $\emph{a priori}$. In fact, it is possible that the opposite scenario, that is of non-equilibrium (non-thermalised) carrier and phonon transport, might actually be more realistic.

\quad The main objective of this work is to develop a dynamic model of electrons and holes in a high density neutral bipolar plasma interacting with longitudinal optical and acoustical phonons, and to use such modelling to analyse the quenching of luminescence in scintillators.

\quad In Section II we discuss the electron-phonon interactions in polar semiconductors, both for acoustic and optical phonons, taking into account screening in a neutral bipolar plasma. We  derive an expression for the ambipolar diffusion coefficient at effective carrier temperatures exceeding room temperature, assuming that at high temperatures, even in extrinsic scintillators with a high activator concentration, the dominant scattering mechanism is due to electron-phonon interactions rather than impurity scattering.  In Section III we derive and solve the balance equations describing the energy exchange between carriers, phonons and the thermal bath.
In Section IV we evaluate the non-proportionality of response for NaI, (the scintillating material, for which most of necessary parameters can be found in the literature) and show that a non-equilibrium model gives results which are consistent with experiment.
Finally in Section V we analyse the implications of our analysis and summarise the significant differences between non-equilibrium and thermalised models.

\section{Electron-phonon interactions in polar semiconductors}

\quad In this Section we summarise textbook results for electron-phonon interactions and present them in a form which is convenient  to use for the analysis of the non-proportional response in scintillators. There are several factors which we must identify as they enter the general expressions. They are: i) screening in a bipolar neutral non-degenerate electron-hole (e-h) plasma and ii) in the high temperature range of interest, $T_e\gg£\hbar\Omega_{LO}$, both the $2k\leq£q_{BZ}$ and $2k\geq£q_{BZ}$ limits, where $k$ is the carrier momentum and $q_{BZ}$ is the phonon quasi-momentum at the Brillouin zone boundary, must be considered.

\subsection{Screening in a non-degenerate, neutral, bipolar electron-hole plasma}

\quad Following the absorption of an X- or gamma-ray photon, the carrier density in the tiny excited volume at the end of the photoelectron track can exceed\cite{Bizarri} $n\geq1.0\cdot£10^{20}$cm$^{-3}$. At such a high density, ignoring the dynamical screening of the interaction between carriers and phonons leads to a serious overestimation of the rates of thermalisation, as well as the rates of momentum relaxation of carriers. In order to take screening into account we will use the Lindhard dielectric response function $\varepsilon(\textbf{q},\omega)$ for a non-degenerate semiconductor, $\displaystyle{\varepsilon(\textbf{q},\omega)=1-\frac{4\pi£e^2}{\varepsilon_0q^2}P(\textbf{q},\omega)}$, where $P(\textbf{q},\omega)$ is the polarisation operator, $\varepsilon_0$ is the static dielectric constant. Calculating  the polarisation operator we write down the expressions for the real $\varepsilon'(q,\omega)$ and imaginary $\varepsilon''(q,\omega)$ parts in the form
\begin{eqnarray}\label{Lindhard}
&&\varepsilon'(x,0)=1+\frac{\sqrt{\pi}}{2x^{3/2}}\left(\frac{\hbar\Omega_p}{T}\right)^2\left[\sqrt{\frac{m_e^*}{m}}\mathrm{erfi}\left(\frac{1}{2}\sqrt{x\frac{m}{m_e^*}}\right)\right.
\times\nonumber\\&&\left.\exp\left(-\frac{x}{4}\frac{m}{m_e^*}\right)+\sqrt{\frac{m_h^*}{m}}\mathrm{erfi}\left(\frac{1}{2}\sqrt{x\frac{m}{m_h^*}}\right)\exp\left(-\frac{x}{4}\frac{m}{m_h^*}\right)\right]\nonumber\\&&
\varepsilon''(x,\omega)=\left\{\begin{array}{c}
                            \displaystyle{\left(\frac{\hbar\Omega_p}{T}\right)^2\frac{\beta}{x^{3/2}}\left[\sqrt{\frac{m_e^*}{m}}
                            \exp\left(-\frac{x}{4}\frac{m}{m_e^*}-\frac{4\beta^2}{x}\frac{m_e^*}{m}\right)\right.}\\
                            \displaystyle{\left.+\sqrt{\frac{m_h^*}{m}}
                            \exp\left(-\frac{x}{4}\frac{m}{m_h^*}-\frac{4\beta^2}{x}\frac{m_h^*}{m}\right)\right],\,\mathrm{LO-phonons}}\\
                           \displaystyle{\left(\frac{\hbar\Omega_p}{T}\right)^2\sqrt{\frac{mv_s^2}{8T}}\frac{1}{x}\left[\sqrt{\frac{m_e^*}{m}}
                            \exp\left(-\frac{x}{4}\frac{m}{m_e^*}\right)\right.}\\
                            \displaystyle{\left.+\sqrt{\frac{m_h^*}{m}}
                            \exp\left(-\frac{x}{4}\frac{m}{m_h^*}\right)\right],\,\mathrm{LA}-\mathrm{phonons}}
                          \end{array}
\right.
\end{eqnarray}
Here $m_{e}^*$ and $m_h^*$ are effective masses for electron and hole respectively, $m$ is free electron mass, $T$ is the carrier temperature (assumed to be equal for electrons and holes), $\displaystyle{\Omega_p=\left(\frac{4\pi£ne^2}{\varepsilon_0m}\right)^{1/2}}$ is the plasma frequency calculated for $m^*_{e,h}=m$. Other notations are $\displaystyle{x=\frac{\hbar^2q^2}{2mT}}$ for the variable,  $\displaystyle{\beta=\frac{\hbar\omega}{4T}}\ll£1$ and $\mathrm{erfi}(x)$ is the imaginary error function. In the expression for $\varepsilon'(x,\omega)$ we took the limiting value  $\varepsilon'(x,0)$ having neglected small terms of the order of $\beta\ll£1$.

\subsection{Phonon-electron interactions}

\quad  The absorption of phonons by a high density non-equilibrium e-h plasma acquires a special significance in relation to non-proportionality in a scintillator. Strong phonon re-absorption may slow down carrier thermalisation. After the absorption of an energetic X or $\gamma$ photon, the e-h density in the cylindrical excitation volume along the photo-electron track is very high, reaching for example a value of $2.0\cdot10^{20}$cm$^{-3}$ in NaI. Such a high density survives for at least a few picoseconds, which is crucial for the formation of a non-proportional response\cite{Li,Bizarri}. The electron scattering times with respect to phonon emission and absorption are much faster (typically on a femtoseconds time scale), while the diffusion of carriers as well as phonon propagation can be considered as slow processes (typical time scale of ps or slower). We must therefore discuss the evolution of the distributions of interacting hot carriers and phonons in much greater detail than was done previously.

\quad The rate of phonon re-absorption by carriers is one of the important parameters of energy exchange between the two systems.
For longitudinal optical phonons (LO-phonons) interacting with carriers through the Fr$\ddot{\mathrm{o}}$lich Hamiltonian, the phonon scattering rate $\Gamma_{\textbf{q}}$ due to interaction with carriers becomes
\begin{eqnarray}\label{Phonon rate}
\Gamma_{\textbf{q}}=-2M_{\textbf{q}}^2\frac{P''(\textbf{q},\omega)}{|\varepsilon(\textbf{q},\omega)|^2}
\end{eqnarray}
Here $\omega=\omega_{\textbf{q}}$ is the dispersion relation for a phonon of a given mode with quasi-momentum $\textbf{q}$, $M_{\textbf{q}}$ is the matrix element describing Fr$\ddot{\mathrm{o}}$lich interaction in ionic crystals, $\displaystyle{M_{\textbf{q}}^2=\frac{2\pi£e^2\hbar\Omega_{LO}}{\varepsilon^*q^2}}$, $e$ is the electron charge, $\displaystyle{\frac{1}{\varepsilon^*}=\left(\frac{1}{\varepsilon_\infty}-
\frac{1}{\varepsilon_0}\right)}$, $\varepsilon_\infty,$ is optical dielectric constant and $\hbar\Omega_{LO}$ is the energy of a longitudinal optical phonon.
Finally $P''(\textbf{q},\omega)$ is the imaginary part of the polarisation operator.

 \quad We consider NaI as an example.  Taking $\varepsilon_0=6.6$, $\varepsilon_\infty=3.0$, $\hbar\Omega_{LO}=$20meV and $m^*_e=0.287m, \,m_h^*=2.397m$ as in Ref\cite{Setyawan}, we plot the graphs of $\Gamma/\Omega_{LO}$ as a function of phonon momentum in units of $\kappa=\sqrt{2m\hbar\Omega_{LO}}/\hbar$. Fig. 1 illustrates the results.
\begin{figure}
  \includegraphics[width=9cm]{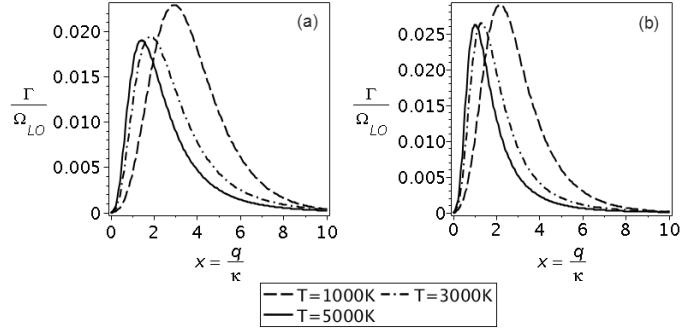}\\
  \caption{Phonon absorption rates by electrons and holes in NaI: (a) $n=2.5\cdot10^{20}$cm$^{-3}$, (b) $n=1.0\cdot10^{20}$cm$^{-3}$ }\label{1}
\end{figure}
It is seen by comparing the results in Fig.1(a) and (b) that screening has a substantial effect, reducing the strength of interaction for higher e-h densities, even though the carrier density in Fig.1(a) is 2.5 times higher than  in Fig.1.(b).

\quad Taking $\Gamma/\Omega_{LO}$=0.01 as a representative value, we obtain $\tau_{ph-e}=\Gamma^{-1}\simeq£3$ps. At the same time, the characteristic time for an energetic ($\sim$1eV) electron or hole to emit a LO-phonon is  estimated to be $\tau_{e-ph}\simeq$5fs using the textbook formulas for unscreened electron-phonon interactions. This is 600 times smaller than $\tau_{ph-e}$. However, electrons and holes are created with high excessive energies above the band gap edges. If we take the commonly accepted relation $E_{e-h}\approx£3E_g$, where $E_{e-h}$ is the mean energy for production of one e-h pair  and $E_g$ is the band gap of the material, then the mean electron (hole) excess energy is $E_g$. This excess energy is rapidly converted into a population of LO-phonons, which cannot escape from the excitation volume because of their low group velocities. Even if the LO-phonon group velocity is taken as high as $1.0\cdot£10^5$cm/s, then before it is re-absorbed by electrons (holes) this phonon would have only passed a distance of $\sim$3nm, less that the cross-section of the excited cylinder. This is one of the main reasons why LO-phonons do not  escape from the excited volume during the whole period of the non-proportionality stage. This results in an effective bottle-necking of energy flow from hot carriers to the thermal bath, slowing down the temperature relaxation of carriers and justifying a model of non-proportionality in scintillators based on non-equilibrium transport.

\quad Another possible channel of relaxing carriers' excess energy is through the anharmonic decay of LO-phonons. The anharmonic rates of LO-phonons decay in NaI are difficult to estimate. In crystals such as GaAs LO-phonons are known to decay in about 5ps, although all decay pathways are open. It is expected that the anharmonic decay of LO-phonon in NaI will be slower. The reason is the large gap between the LO-energy and maximum energy of LA-phonons, $\hbar\Omega_{LO}\geq£2\hbar\Omega_{LA}$, prohibiting the possibility of LO-phonon decay into two acoustical phonons. For other scintillators, for example CsI, the same gap is smaller, thus  allowing three-phonon anharmonic decay into acoustical phonons. However,  most interaction channels  will allow energy from LO-phonons to be transferred predominantly to TO-phonons, which have almost zero group velocities. Thus energy can escape from the excited volume only through the generation of acoustic phonons, being either the products of direct emission by thermalising electrons (holes) or of anharmonic decays of LO- and TO-phonons. In the first instance, the rate of energy loss due to emission of acoustical LA-phonons is typically one to two orders of magnitude smaller than that for LO-emission. In the second case, the energy released directly in the form of LA and TA phonons is only a small fraction of the LO-energy, and the TO-phonons contributing to the energy flow bottle-neck.

\quad Summarising, it is likely that over the duration of the non-proportionality stage the energy of the initial photon within the excited volume will be  shared between  hot carriers and LO-phonons. A larger fraction of this energy will be locked in the LO-phonon distribution. Indeed, in the fast process of LO-phonon emission, the excess energy of each carrier converts into the LO-phonon distribution. The excess energy released in the form of LO-phonons may be estimated as $E_g-T$ per carrier. For NaI $E_g$=5.8eV, so taking $T\leq5000$K implies that 5eV per carrier will be converted into approximately 2500 LO-phonons (20meV each). The actual numbers will be somewhat less if the parallel emission of acoustical phonons and possibly carrier-carrier  interactions are taken into account. Even with 1000 LO-phonons per carrier within the excited volume the following scenario is possible.
Within a 5fs period  each carrier is likely to emit an LO-phonon; on the other hand the likelyhood of absorption of a single LO-phonon is 600 times smaller ($\tau_{ph-e}$=3ps). However, the probability of absorbing one of 1000 available LO-phonons per carrier is 1000 times larger. This qualitative reasoning shows that over the non-proportionality stage there will be balance established between the carriers and LO-phonons. In a quasi-stationary state, the energy loss from hot electrons and holes will be balanced by phonon re-absorption. The hot carrier temperature, $T$, will be determined by the energy escape rate from the excitation volume through the emission of acoustic phonons and the diffusion of carriers leaving hot phonons behind.

\subsection{Scattering of electrons and holes by LO- and LA-phonons}

\quad In this section we will analyse the rates of momentum relaxation of carriers due to Fr$\mathrm{\ddot{o}}$lich interaction with LO-phonons, and deformation potential interaction with LA-phonons. General expressions can be found in textbooks\cite{Anselm}. We write down the standard expression neglecting dispersion for LO-phonons and assuming a linear dispersion relation for LA-phonons. We additionally take into account dynamic screening, and make no assumptions about the magnitude of the ratio $2k/q_{BZ}$.
\begin{eqnarray}\label{momentum rate}
&&\frac{1}{\tau_{e(h)}(x)}=\frac{e^2}{4\varepsilon^*}\frac{(2mT)^{1/2}}{\hbar^{2}x^{3/2}}\frac{m^*_{e(h)}}{m}\int_0^{\psi(x,T)}\frac{\mathrm{d}x'}{|\varepsilon(\textbf{q},\omega)|^2}
\times\nonumber\\&&\left(1+\frac{\varepsilon^*\Lambda_{e(h)}^2}{2\pi\rho£v_s^2}\frac{mT}{\hbar^2£e^2}x'\right)
\end{eqnarray}
where the upper integration limit is $\displaystyle{\psi(x,T)=4x\Theta\left(\frac{\hbar^2q_{BZ}^2}{2mT}-4x\right)+\frac{\hbar^2q_{BZ}^2}{2mT}}\Theta\left(4x-\frac{\hbar^2q_{BZ}^2}{2mT}\right)$, $\Lambda_{e(h)}$ is the deformation potential constant for electrons (holes), $\rho$ the crystal density and $v_s$ the longitudinal sound velocity.

\quad Taking into account only carrier interaction with LO- and LA-phonons we obtain the mobility of electrons(holes) $\mu_{e(h)} $ in the form
\begin{eqnarray}\label{mobility}
&&\mu_{e(h)}=\frac{16}{3\sqrt{2\pi}}\frac{\varepsilon^*\hbar^2}{m^{3/2}T^{1/2}e}\left(\frac{m}{m^*_{e(h)}}\right)^{9/2}\left\{\int_0^{\hbar^2q_{BZ}^2/8mT}\mathrm{d}x\right.\nonumber\\&&
x^3\exp\left(-x\frac{m}{m^*_{e(h)}}\right)\left[\int_0^{4x}\frac{\mathrm{d}x'}{|\varepsilon(x',\omega)|^2}\left(1+\frac{\varepsilon^*\Lambda_{e(h)}^2}{2\pi\rho£v_s^2}\frac{mT}{\hbar^2£e^2}x'\right)\right]^{-1}+
\nonumber\\&&\int_0^{\hbar^2q_{BZ}^2/8mT}\mathrm{d}x
x^3\exp\left(-x\frac{m}{m^*_{e(h)}}\right) \times \nonumber\\&&\left.\left[\int_0^{\hbar^2q^2_{BZ}/2mT}\frac{\mathrm{d}x'}{|\varepsilon(x',\omega)|^2}\left(1+\frac{\varepsilon^*\Lambda_{e(h)}^2}{2\pi\rho£v_s^2}\frac{mT}{\hbar^2£e^2}x'\right)\right]^{-1}\right\}
\end{eqnarray}

\section{ Dynamics of hot carriers and phonons}

In this section, we derive the balance equations for electrons and  holes interacting with LO- and LA-phonons. Electrons and holes are assumed to be at the same quasi-equilibrium temperature at all times. This implies strong carrier-carrier (e-e, e-h and h-h) interactions.  Therefore we may consider electron and holes as a single sub-system with a heat capacity equal to the total heat capacity of electrons and holes. The heat conductance describing the exchange of energy between carriers and phonons is a sum of individual (different) heat conductances, linking electrons and holes with the respective phonon sub-systems.  Fig. 2 shows schematically the energy flow from the excitation volume into the thermal bath. Fig. 2(a) is more general, allowing energy relaxation of the system carriers+LO-phonons, where it was originally deposited, through direct emissions of LA-phonons and their subsequent escape (ballistic or diffusive) from the excited volume. Another relaxation channel is through the anharmonic decays of LO-phonons into their decay products, namely TO-, TA- and LA-phonons and subsequent escape of acoustical phonons from the excitation volume. Overall this balance scheme, although realistic, cannot be modelled accurately without a knowledge of the details of anharmonic interactions. For this reason we consider the simplified scheme depicted in Fig. 2(b) where we incorporate all escape routes into LA-and LO-phonon decays, introducing an effective decay power $P_{LA,decay}(T_{LA})=\gamma_{LA}£C_{LA}(T_{LA}-T_0)$ and $P_{LO,decay}(T_{LO})=\gamma_{LO}£C_{LO}(T_{LA}-T_0)$, where $\gamma_{LA(LO)}$ are the effective decay rates, and $C_{LA(LO)}$ is the specific heat of LA- or LO-phonons.
\begin{figure}
  \includegraphics[width=8cm]{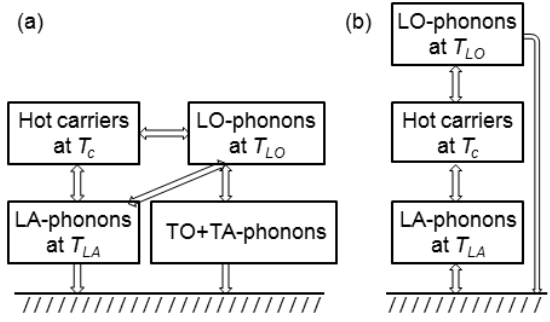}\\
  \caption{Schematic of energy exchange channels: (a) general scheme, (b) simplified scheme}\label{3}
\end{figure}

\quad The balance equations describing energy exchange as depicted in Fig.2(b) have the form
\begin{eqnarray}\label{balance}
&&\frac{dE_c}{dt}=P_{c-LO}(T_e,T_{LO})+P_{c-LA}(T_e,T_{LA})+E_c(0)\delta(t)\nonumber\\&&
\frac{dE_{LO}}{dt}=-P_{c-LO}(T_e,T_{LO})-P_{LO,decay}(T_{LO},T_0)+\nonumber\\&&£E_{LO}(0)\delta(t)\nonumber\\&&
\frac{dE_{LA}}{dt}=-P_{c-LA}(T_e,T_{LA})-P_{LA,decay}(T_{LA},T_0)
\end{eqnarray}
Here $E_c=E_e+E_h,\,E_{LO},\,E_{LA}$ and $T_e,\,T_{LO},\,T_{LA},\,T_0$ are energy densities and temperatures of carriers, LO-phonons and LA-phonons respectively, $T_0$ is the thermal bath temperature. In the right hand side of expression (\ref{balance}) $P_{c-LO}(T_e,T_{LO})$ and $P_{c-LA}(T_e,T_{LA})$ are the power densities dissipated by carriers into LO- and LA-phonons respectively, $P_{c-LO}(T_e,T_{LO})=P_{e-LO}(T_e,T_{LO})+P_{h-LO}(T_e,T_{LO})$ and $P_{c-LA}(T_e,T_{LA})=P_{e-LA}(T_e,T_{LA})+P_{h-LA}(T_e,T_{LA})$. Finally, $P_{LO,decay}(T_{LO},T_0)$ describes the dissipated power density which escapes from the excited volume via the anharmonic decay products of LO-phonons, and similarly $P_{LA,decay}(T_{LA},T_0)$ has been introduced to describe the combined effect of LA-phonons being generated as a result of anharmonic decay of LO-phonons and LA-phonons and  escaping from the excitation volume.
Also in (\ref{balance}), $E_c(0)$ and $E_{LO}(0)$ are energy densities deposited into electrons and holes and LO-phonons respectively, at $t=0$, so that if $E_0$ is the photon energy, then $E_0=E_c(0)+E_{LO}(0)$, neglecting any direct emission of acoustic phonons in the process of the initial generation of electron-hole pairs.

\subsection{Power dissipation from carriers to phonons}

\quad The general expression for the energy loss from the electron system due to interaction with phonons has the form\cite{Allen}
\begin{eqnarray}\label{Pe-LO}
&&P_{e-ph}(T_e,T_{ph})=-\frac{4\pi}{\hbar}\sum_{\textbf{k},\textbf{q}}\frac{\epsilon_\textbf{k}M_{\textbf{q}}^2}{|\varepsilon(\textbf{q},\omega)|^2}\left\{\left[(N_{\textbf{q}}+1)\times\right.\right.\nonumber\\&&\left.
f_{\textbf{k}}(1-f_{\textbf{k}-\textbf{q}})-N_{\textbf{q}}f_{\textbf{k}-\textbf{q}}(1-f_{\textbf{k}})\right]\delta\left(\epsilon_{\textbf{k}}-\epsilon_{\textbf{k}-\textbf{q}}-\hbar\omega_{\textbf{q}}\right)-\nonumber\\&&
\left[(N_{\textbf{q}}f_{\textbf{k}+\textbf{q}}(1-f_{\textbf{k}})-N_{\textbf{q}}f_{\textbf{k}}(1-f_{\textbf{k}+\textbf{q}})\right]\times\nonumber\\&&\delta\left(\epsilon_{\textbf{k}}-\epsilon_{\textbf{k}+\textbf{q}}-\hbar\omega_{\textbf{q}}\right)
\end{eqnarray}
where $f_{\textbf{k}}$ is the electron distribution function,  $N_{\textbf{q}}$ is phonon distribution function, and $\epsilon_{\textbf{k}}$ is the electron dispersion relation. A similar expression can be written for holes. We will incorporate the additive hole contributions  later directly in the balance equations.
 The expression for power dissipation in phonon collisions with the electron gas is
\begin{eqnarray}\label{Pphe}
&&P_{ph-e}(T_e,T_{ph})=-\frac{4\pi}{\hbar}\sum_{\textbf{k},\textbf{q}}\frac{\hbar\omega£M^2_{\textbf{q}}}{|\varepsilon(\textbf{q},\omega)|^2}
f_{\textbf{k}}\left[(1-f_{\textbf{k}+\textbf{q}})\right.\times\nonumber\\&&N_{\textbf{q}}\delta(\epsilon_{\textbf{k}}-\epsilon_{\textbf{k}+\textbf{q}}+
\hbar\omega_{\textbf{q}})-(N_{\textbf{q}}+1)(1-f_{\textbf{k}-\textbf{q}})\times\nonumber\\&&\left.
\delta(\epsilon_{\textbf{k}}-\epsilon_{\textbf{k}-\textbf{q}}-
\hbar\omega_{\textbf{q}})\right]
\end{eqnarray}
Here $M_{\textbf{q}}$ is the appropriate matrix element for the electron-phonon interaction
 \begin{eqnarray}\label{coupling}
 \left\{\begin{array}{c}
   \displaystyle{M_{\textbf{q}}^2=\frac{2\pi£e^2\hbar\Omega_{LO}}{\varepsilon^*q^2}}\, \mathrm{for\,polar\, coupling} \\
   \displaystyle{M_{\textbf{q}}^2=\frac{1}{2
 }\frac{\hbar\Lambda_e^2£q}{\rho£v_s}}\, \mathrm{for\, deformation\, potential\, coupling}
 \end{array}\right.
 \end{eqnarray}

\quad The total energy of interacting carriers and phonons is conserved within the system leading to $P_{e-ph}(T_e,T_{ph})+P_{ph-e}(T_e,T_{ph})=0$. This conservation relation has already been used in the balance equations (\ref{balance}).

\quad Instead of evaluating the dissipated power $P_{e-ph}(T_e,T_{ph})$ using (\ref{Pe-LO}) we write it as -$P_{ph-e}(T_e,T_{ph})$ using expression (\ref{Pphe}).
For non-degenerate electrons this becomes
\begin{eqnarray}\label{PPHE1}
&&P_{ph-e}(T_e,T_{ph})=4\pi\hbar\left(\frac{1}{T_{ph}}-\frac{1}{T_{e}}\right)\exp\left(\frac{\mu}{T_e}\right)\times\nonumber\\&&\sum_{\textbf{k},\textbf{q}}
\frac{\omega^2£M^2_{\textbf{q}}N_{\textbf{q}}}{|\varepsilon(\textbf{q},\omega)|^2}\exp\left(-\frac{\epsilon_{\textbf{k}}}{T_e}\right)
\times\nonumber\\&&\delta(\epsilon_{\textbf{k}}-\epsilon_{\textbf{k}+\textbf{q}}+
\hbar\omega_{\textbf{q}})
\end{eqnarray}
where $\mu$ is the electron chemical potential.
Calculating the integral over $\textbf{k}$  we obtain
\begin{eqnarray}
&&\sum_{\textbf{k}}
\exp\left(-\frac{\epsilon_{\textbf{k}}}{T_e}\right)
\delta(\epsilon_{\textbf{k}}-\epsilon_{\textbf{k}+\textbf{q}}+
\hbar\omega_{\textbf{q}})=\nonumber\\&&\frac{m^{*2}T_e}{4\pi^2\hbar^4q}\exp\left(-\frac{\hbar^2q^2}{8m^*T_e}-\frac{m^*\omega^2}{2q^2T_e}
+\frac{\hbar\omega}{4T_e}\right)
\end{eqnarray}

\quad Substituting this result into (\ref{PPHE1}) yields $P_{e-LO}(T_e,T_{LO})$. Adding a similar hole contribution we obtain
\begin{eqnarray}\label{PLOe}
&&P_{c-LO}(T_e,T_{LO})=-\frac{\Omega_p^2\kappa^3\beta^{3/2}\varepsilon_0}{\Omega_{LO}\varepsilon^*}f(T_e)\left(T_e-T_{LO}\right)
\end{eqnarray}
where we have introduced the function $f(T_e)$ according to
\begin{eqnarray}
 && f(T_e)=\frac{1}{2\pi^{1/2}}\int_0^{\hbar^2q_{BZ}^2/2mT_e}\frac{\mathrm{d}x}{x}\frac{1}{|\varepsilon(x,0)|^2}\times\nonumber\\&&
 \left[\left(\frac{m_e^*}{m}\right)^{1/2}\exp\left(-\frac{x}{4}\frac{m}{m_e^*}-\frac{4\beta^2}{x}\frac{m_e^*}{m}\right)+
 \left(\frac{m_h^*}{m}\right)^{1/2}\right.\times\nonumber\\&&\left.\exp\left(-\frac{x}{4}\frac{m}{m_h^*}-\frac{4\beta^2}{x}\frac{m_h^*}{m}\right)\right]
  \end{eqnarray}

\quad Similarly for power dissipated into LA-phonons we obtain
\begin{eqnarray}\label{1LA}
&&P_{c-LA}(T_e,T_{LA})=-\frac{4}{\pi^{3/2}}\frac{nm\Lambda^2}{\rho\hbar£T_e}\left(\frac{2mT_e}{\hbar^2}\right)^{3/2}\times\nonumber\\&&g(T_e)\left(T_e-T_{LA}\right)
\end{eqnarray}
where
\begin{eqnarray}
 &&g(T_e)=\frac{1}{16}\int_0^{\hbar^2q_{BZ}^2/2mT_e}\frac{\mathrm{d}xx}{|\varepsilon(x,0)|^2}\left[\left(\frac{m_e^*}{m}\right)^{1/2}\frac{\Lambda^2_e}{\Lambda^2}\right.\times\nonumber\\
 &&\left.\exp\left(-\frac{x}{4}\frac{m}{m_e^*}\right)+\left(\frac{m_h^*}{m}\right)^{1/2}\frac{\Lambda^2_h}{\Lambda^2}\exp\left(-\frac{x}{4}\frac{m}{m_h^*}\right)\right]
\end{eqnarray}
where $\Lambda$ is a mean deformation potential constant.

 \subsection{Solving the balance equations}

 \quad The system of balance equations (\ref{balance}) has three unknown variables, namely $T_{e} (t), T_{LO}(t)$ and $T_{LA}(t)$ (the three temperature model). It is convenient to introduce dimensionless variables, $x(t)=T_e(t)/T(0)$, $y(t)=T_{LO}(t)/T(0)$ and $z(t)=T_{LA}(t)/T(0)$, where $T(0)$ is the initial temperature of carriers and LO-phonons, which is established after the generation of e-h pairs and quasi-equilibration within the carrier-LO-phonon system.  We may also introduce a dimentionless time $t$ defined in special units $\displaystyle{t_0=\frac{3\pi^{3/2}}{2^{7/2}}\frac{\rho\hbar^4}{m^{5/2}\Lambda^2(T(0))^{1/2}}}$.
 The initial temperature $T(0)$ can be found from
 \begin{eqnarray}\label{ini}
 3nT(0)+N_cT(0)=2nE_g
 \end{eqnarray}
 where $N_c$ is number of elementary cells per unit volume. This condition implies that the excess energy density, which will be released prior the recombination of the generated e-h pairs, is equal to locally thermalized (within carriers and LO-phonons) energy. At high temperature, $3nT(0)$ is the total energy of locally thermalised carriers in the excited volume, and $N_cT(0)$ is the energy density deposited into a single phonon mode (LO-phonon).

   \quad Re-writing the balance equations in dimensionless units we obtain
  \begin{eqnarray}\label{2T-balance}
&&\frac{dx}{dt}=-\frac{2\pi^{5/2}\rho£e^2\hbar^2\beta^2(0)}{\varepsilon^*m^2\Lambda^2} x^{-3/2}f(x)(x-y)\nonumber\\&&-x^{1/2}g(x)(x-z)+\delta(t)\nonumber\\&&
\frac{dy}{dt}=\frac{n}{N_c}\frac{6\pi^{5/2}\rho£e^2\hbar^2\beta^2(0)}{\varepsilon^*m^2\Lambda^2} x^{-3/2}f(x)(x-y)-\nonumber\\&&
-\gamma_{LO}t_0\left(y-\frac{T_0}{T(0)}\right)+\delta(t)\nonumber\\&&
\frac{dz}{dt}=\frac{3n}{N_c}x^{1/2}g(x)(x-z)-\gamma_{LA}t_0\left(z-\frac{T_0}{T(0)}\right)
 \end{eqnarray}
 Eq.(\ref{2T-balance}) can be solved numerically provided that the main material parameters are known. We will test our model for three different scintillator crystals: NaI, CsI and ZnSe. For all these crystals most of important material parameters can be either found in the literature or estimated. Where we could not find relevant data we show parametric graphs, allowing the unknown parameter to vary within an expected range.

 \subsection{NaI, CsI, ZnSe: solution of balance equations}

 \quad The estimate for the initial temperature T(0) in NaI from (\ref{ini}) is $T(0)=4440$K for $n=2.5\cdot£10^{20}$cm$^{-3}$. For $n=1.0\cdot£10^{20}$cm$^{-3}$ we obtain $T(0)=2050$K. In both cases, we have $T(0)\gg£T_0$.
Using $\rho=3.67$g$\cdot$cm$^{-3}$ we obtain the  characteristic time $\displaystyle{t_0=8.9\cdot£10^{-9}\left(\frac{1\mathrm{eV}}{\Lambda}\right)^2\left(\frac{1\mathrm{K}}{T(0)}\right)^{1/2}}$s. For $\Lambda=10$eV we obtain $t_0=1.34$ps for $n=2.5\cdot£10^{20}$cm$^{-3}$ and $t_0=1.97$ps for $n=1.0\cdot£10^{20}$cm$^{-3}$.
Figs. 3 and 4 show the solution of balance equations in NaI for the two values of on-axis e-h density, different values of phonon decay rates and two values of deformation potential constant (taking the same value for electrons and holes) falling into typical value ranges for known semiconductors, $\Lambda=5$eV and $\Lambda=10$eV.
\begin{figure}
  \includegraphics[width=8cm]{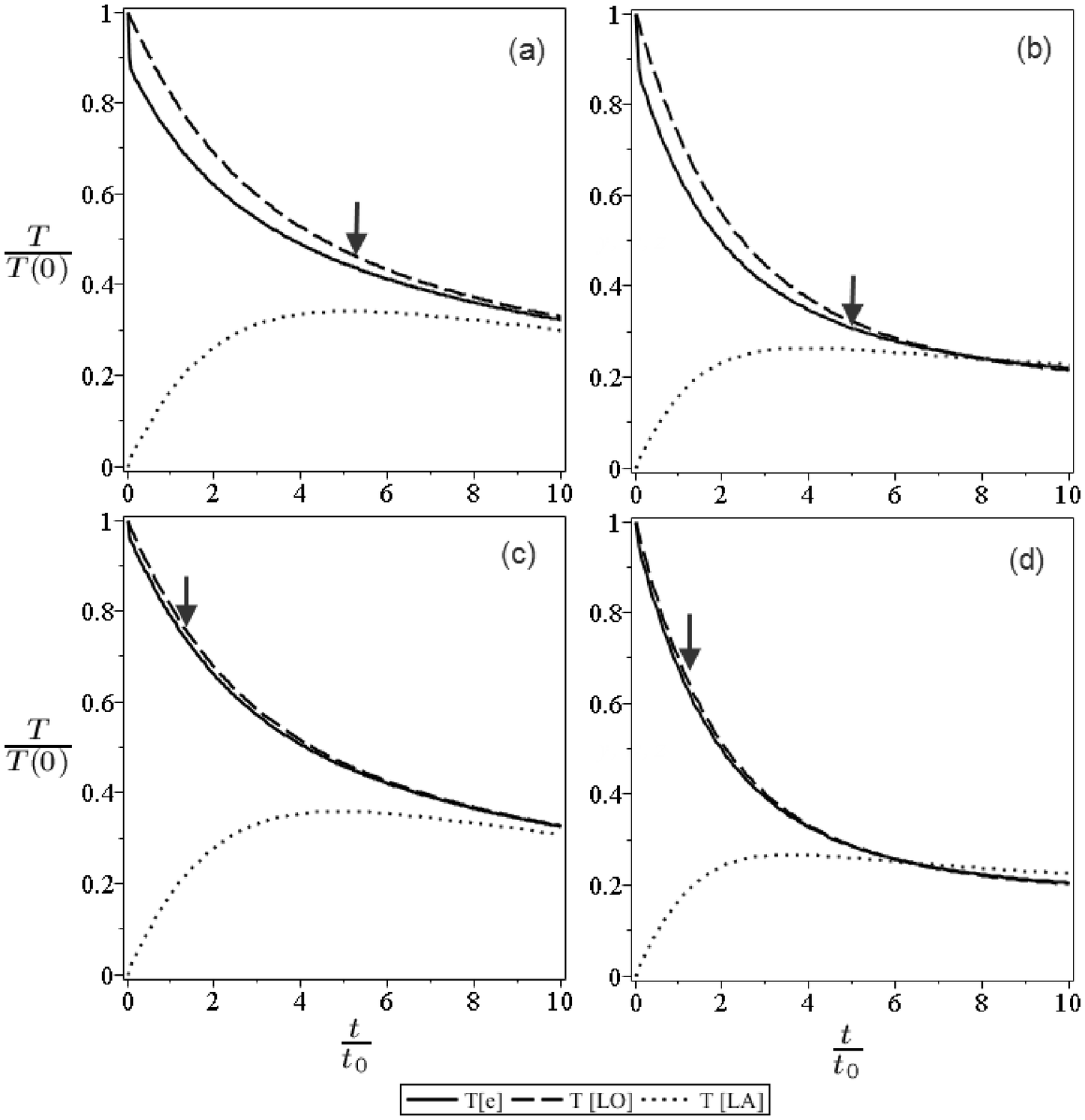}\\
  \caption{Relaxation of temperature of carriers and phonons in NaI, $n=1.0\cdot10^{20}$cm$^{-3}$: (a) - $\Lambda=10$eV,\, $\gamma_{LO}t_0=\gamma_{LA}t_0=0.1$; (b) - $\Lambda=10$eV,\, $\gamma_{LO}t_0=0.1,\, \gamma_{LA}t_0=1$;
  (c) - $\Lambda=5$eV,\, $\gamma_{LO}t_0=\gamma_{LA}t_0=0.1$; (d) - $\Lambda=5$eV,\, $\gamma_{LO}t_0=0.1, \,\gamma_{LA}t_0=1$. Arrows indicate $t$=10ps}\label{3}
\end{figure}
\begin{figure}
  \includegraphics[width=8cm]{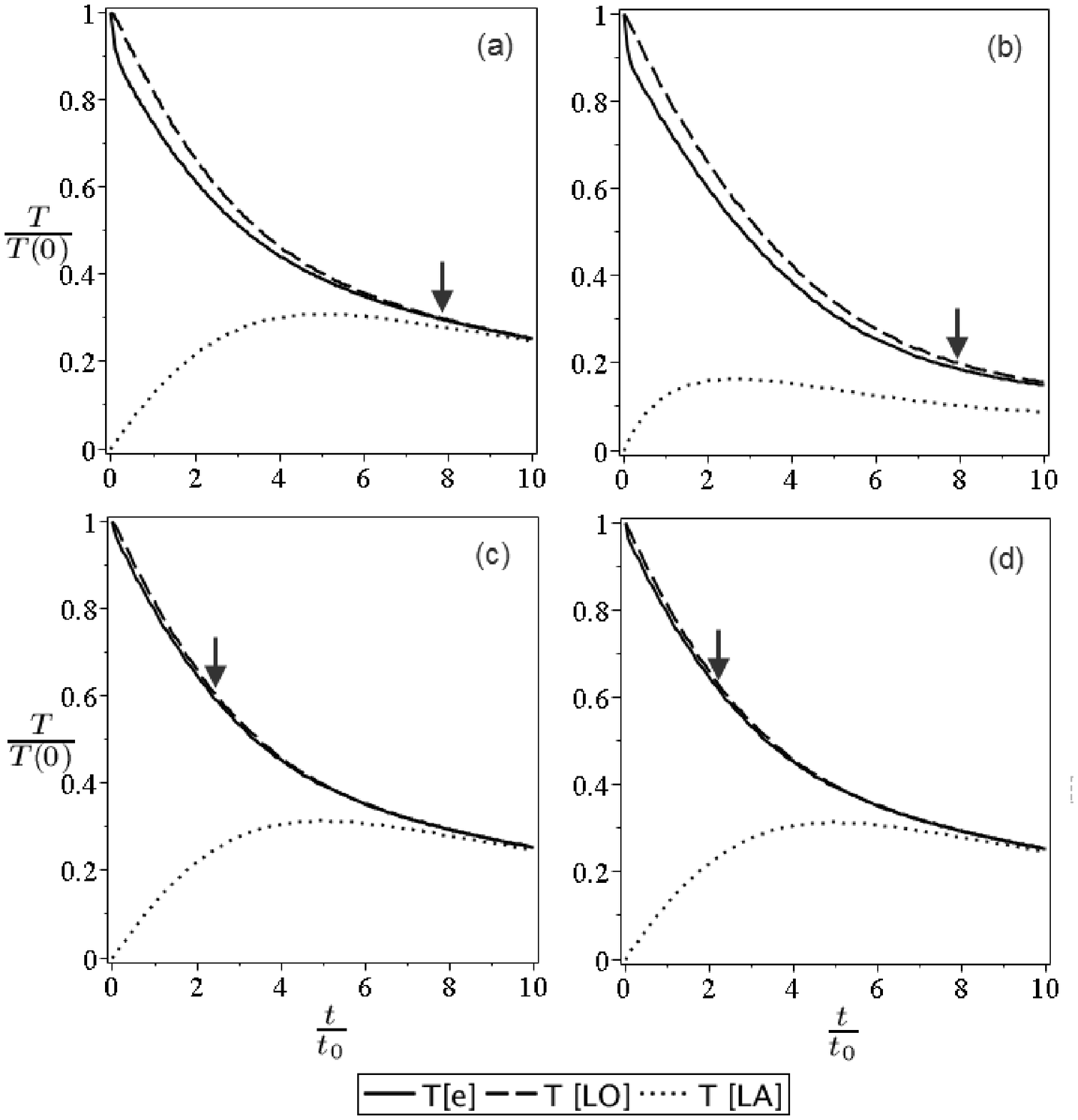}\\
  \caption{Relaxation of temperature of carriers and phonons in NaI, $n=2.5\cdot10^{20}$cm$^{-3}$: (a) - $\Lambda=10$eV,\, $\gamma_{LO}t_0=\gamma_{LA}t_0=0.1$; (b) - $\Lambda=10$eV,\, $\gamma_{LO}t_0=0.1,\, \gamma_{LA}t_0=1$;
  (c) - $\Lambda=5$eV,\, $\gamma_{LO}t_0=\gamma_{LA}t_0=0.1$; (d) - $\Lambda=5$eV,\, $\gamma_{LO}t_0=0.1, \,\gamma_{LA}t_0=1$. Arrows indicate $t$=10ps}\label{4}
\end{figure}
It is seen from Figs.3-4 that within a 10ps interval the decreasing carrier and LO-phonon temperature depends, as expected,  on the effectiveness of LA-phonon emission through the value of $\Lambda$. For $n=1.0\cdot10^{20}$cm$^{-3}$ this means that the temperature decreases by $\sim£50\%$ for $\Lambda=10$eV and $\sim£20\%$ for $\Lambda=5$eV. For $n=2.5\cdot10^{20}$cm$^{-3}$ the decrease in temperature over 10ps interval is larger, $60\%$ and $40\%$ respectively, mostly because the initial temperature is higher, and the LA-scattering dependence of temperature is stronger than for LO-scattering.

\quad Finally, it may look unrealistic to use the concept of a peak temperature of the distribution of LO-phonons as high as 4440K, exceeding the melting temperature (961K) of NaI .
However, the distribution of LO-phonons created by the energy relaxation of hot carriers in a tiny excited volume is transient and far from equilibrium, implying that lattice vibrations of only one mode are present, whereas all other modes remain undisturbed. The heat capacity of the LO-mode is only 1/6 of the overall heat capacity, implying that the energy in this vibrational state or squared amplitude of vibrations would be 1/6th of hypothetical equilibrium value at 4440K. In this way the amplitudes of atomic vibrations with only LO-mode excited to an effective temperature of 4440K are equivalent to the amplitudes of all vibrations at an equilibrium situation at 1/6 of 4440K or 740K, which is below the melting point.

 \quad The estimate for the initial temperature, T(0), in CsI from (\ref{ini}) is $T(0)=2905$K for $n=1.0\cdot£10^{20}$cm$^{-3}$.
Using $\rho=4.51$g$\cdot$cm$^{-3}$ and $\Lambda=10$eV we obtain $t_0=2.0$ps for $n=1.0\cdot£10^{20}$cm$^{-3}$.
Fig. 5 illustrates the solution of the balance equations in CsI.
\begin{figure}
  \includegraphics[width=8cm]{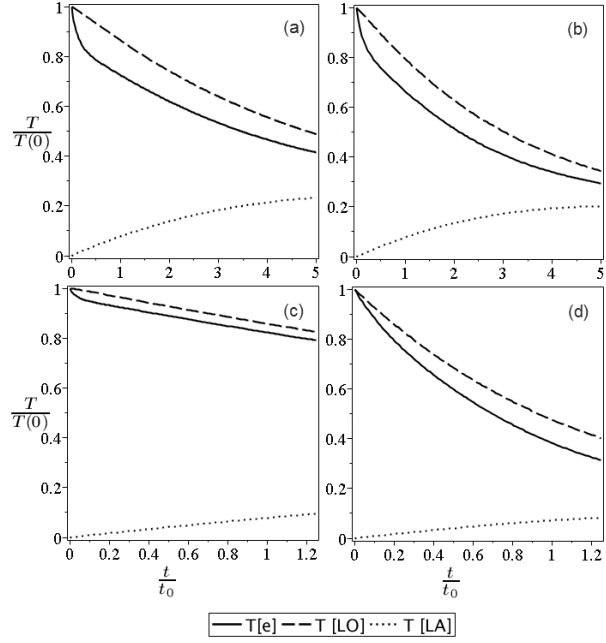}\\
  \caption{Relaxation of temperature of carriers and phonons in CsI, $n=1.0\cdot10^{20}$cm$^{-3}$: (a) - $\Lambda=10$eV,\, $\gamma_{LO}t_0=\gamma_{LA}t_0=0.1$; (b) - $\Lambda=10$eV,\, $\gamma_{LO}t_0=0.2,\, \gamma_{LA}t_0=0.1$;
  (c) - $\Lambda=5$eV,\, $\gamma_{LO}t_0=\gamma_{LA}t_0=0.1$; (d) - $\Lambda=5$eV,\, $\gamma_{LO}t_0=0.85, \,\gamma_{LA}t_0=1$}\label{3}
\end{figure}
The limits along the horizontal axis were chosen to cover the interval 0-10ps.
In Figs.5 (b) and (d) we modelled the effect of anharmonic decay of the LO-phonon, assuming the LO-phonon decay time to be 10ps, falling within a typical range for the spontaneous decay of optical phonons of similar energy in semiconductors.

\quad For ZnSe, using the material parameters, we obtain $T(0)$=880K for $n=1.0\cdot10^{20}$cm$^{-3}$ and $T(0)$=1700K for $n=2.5\cdot10^{20}$cm$^{-3}$. These smaller values reflect the smaller band gap in ZnSe, $E_g=2.82$eV. The smaller value of the pair-production energy may well result in a higher on-axis density of e-h pairs within the cylindrical excited volume; thus the second estimate for e-h pairs density looks preferable. Nonetheless, we will also discuss the results for a smaller density, where the initial state is closer to equilibrium.
Fig. 6 illustrates the solution of the balance equations in ZnSe for $n=2.5\cdot10^{20}$cm$^{-3}$. The limits along the horizontal axis were chosen to cover the interval 0-10ps ($t_0=3.1$ps for $\Lambda=10$eV).
\begin{figure}
  \includegraphics[width=8cm]{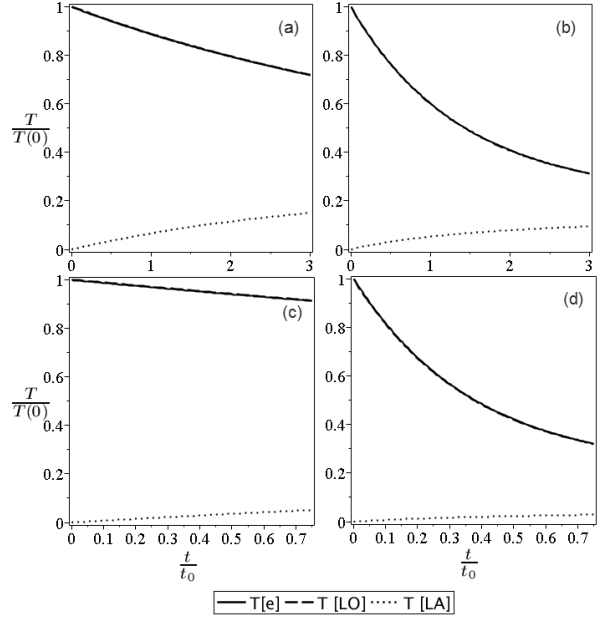}\\
  \caption{Relaxation of temperature of carriers and phonons in ZnSe, $n=2.5\cdot10^{20}$cm$^{-3}$: (a) - $\Lambda=10$eV,\, $\gamma_{LO}t_0=\gamma_{LA}t_0=0.1$; (b) - $\Lambda=10$eV,\, $\gamma_{LO}t_0=0.7,\, \gamma_{LA}t_0=0.1$;
  (c) - $\Lambda=5$eV,\, $\gamma_{LO}t_0=\gamma_{LA}t_0=0.1$; (d) - $\Lambda=5$eV,\, $\gamma_{LO}t_0=2.8, \,\gamma_{LA}t_0=1$.}\label{6}
\end{figure}
 In Figs.6(b) and (c) we use
$\gamma_{LO}^{-1}$=5ps, equal to that measured in GaAs, which has a similar phonon spectrum and close values for third order elastic constants. Also the data from literature\cite{Bairamov} give $\Lambda$ in the range 6.8-10.2eV.
As seen from Fig.6, even with a reasonably strong anharmonic decay over a 10ps interval, the carrier temperature only falls  to $\sim$500K, remaining significantly higher that room temperature. For a smaller density $n=1.0\cdot10^{20}$cm$^{-3}$ without anharmonic decay  the carrier temperature changes within $\leq$10$\%$ of $T(0)$; with the estimated anharmonic decay it decreases down to 360K, implying that the carrier transport is not completely thermalised.

\section{Non-proportional light yield in $\mathrm{NaI}$}

\quad In this section we will discuss implications of the non-equilibrium transport model, checking its predictions versus the thermalised transport model of Li $\emph{et al}$\cite{Li}.

\quad Following Payne $\emph{et al}$ we write down the scintillator yield in the form\cite{payne1,payne2}
\begin{eqnarray}\label{payne}
\eta_{PE}=N_0\eta_{CAP}\eta_{PH}
\end{eqnarray}
where $N_0$ is the number of generated e-h pars, $\eta_{CAP}$ is the efficiency by which
the carriers' energy is captured by the activators, and $\eta_{PH}$ is
the efficiency of light emission by activators. We will model $\eta_{CAP}$ using Birks and Onsager mechanisms\cite{payne1,payne2}.
\begin{eqnarray}\label{etaCAP}
\eta_{CAP}\sim£\frac{1-\eta_{e/h}\exp(-\frac{r_{Ons}}{r_{e/h}})}{1+\frac{dE/dx}{dE/dx|_{Birks}}}
\end{eqnarray}
Here $r_{Ons}$ is the Onsager radius defined according to $\displaystyle{\frac{e^2}{\varepsilon_0r_{Ons}}=k_BT}$, $r_{e/h}$ is the electron-hole separation, $\eta_{e/h}$ is the fraction of free electrons and holes formed
during the cascade, $dE/dx$ is the electron energy deposition rate, and $dE/dx|_{Birks}$ is an empirical fitting parameter relating
to the strength of the exciton-exciton annihilation mechanism.
The right hand side of (\ref{etaCAP}) is the scintillator light yield non-proportionality factor, which is a function of photon energy $E_0$ and carrier temperature $T$, which we will designate as $\varrho(n)$. For a proportional scintillator $\varrho(n)=1$.

\quad Instead of the empirical Birks relation we will use a model of bimolecular quenching due to dipole-dipole F$\mathrm{\ddot{o}}$rster
transfer and introduce a fraction of carriers surviving the non-linear quenching, $1-QF$, where QF is the quenching factor\cite{Li}. Thus, we obtain the local non-proportionality factor for a section of the track with carrier density $n$ in the form
\begin{eqnarray}\label{nf}
\varrho(n)=(1-QF)\left[1-\eta_{e/h}\exp\left({-\frac{r_{Ons}}{r_{e/h}}}\right)\right]=\varrho_1(n)\varrho_2(n)
\end{eqnarray}
where $\varrho_1(n)=1-QF$ and $\varrho_2(n)=\left[1-\eta_{e/h}\exp\left(-\displaystyle{\frac{r_{Ons}}{r_{e/h}}}\right)\right]$.
To obtain the integral non-proportionality factor we must take an integral of (\ref{nf}) along  the trajectory of photoelectron.
 If the quenching factor $QF\leq1/2$ we may evaluate it analytically by integrating the quadratic form of linearised solutions to the two-dimensional diffusion equation with cylindrical symmetry of the excited volume and iterating up to the second order terms in the on-axis carrier density. As a result we obtain
 \begin{eqnarray}\label{QF}
&& \varrho_1(n)=1-\overline{k_2}\sqrt{\tau}n\frac{\arctan\left(2\sqrt{D_a\tau/a^2}\right)}{2\sqrt{D_a\tau/a^2}}+\nonumber\\&&\frac{\overline{k_2^2}\tau£n^2}{2\sqrt{D_a\tau/a^2}}
\left\{\frac{1}{2}\left[\arctan\left(2\sqrt{D_a\tau/a^2}\right)\right]^2+\right.\nonumber\\&&\left.\int_0^{2\sqrt{D_a\tau/a^2}}\mathrm{d}s\frac{\mathrm{arctanh}\displaystyle{\left(\frac{s}{\sqrt{3+4s^2}}\right)}}{(1+s^2)\sqrt{3+4s^2}}\right\}
 \end{eqnarray}
 where  $n$ is the on-axis density of e-h pairs, $D_a$ is the ambipolar diffusion coefficient in a bipolar plasma of electrons and holes, $a$ is the radius of  the excited cylinder, and $\tau$ is the characteristic time, equal to the duration of an interval over which non-proportionality of response develops. The coefficient $k_2(t)=\overline{k_2}/\sqrt{t}$ is the bimolecular quenching rate parameter due to dipole-dipole F$\mathrm{\ddot{o}}$rster
transfer. This has been measured in CsI\cite{Williams} to be  $k_2(t)\sqrt{t}=2.4\cdot10^{-15}$cm$^3$s$^{-1/2}$. Following Li $\emph{et al}$\cite{Li} we will use the same value to estimate exciton loss in NaI.
For the ambipolar diffusion coefficient we  take $\displaystyle{D_a=\frac{\mu_eD_h+\mu_hD_e}{\mu_e+\mu_h}=\frac{\mu_e\mu_h}{\mu_e+\mu_h}\frac{T_e+T_h}{e}}$.
For $T_e=T_h$ we obtain $\displaystyle{D_a=\frac{2\mu_e\mu_h}{\mu_e+\mu_h}\frac{T_e}{e}}$.

\quad Fig.7 shows the results of calculation of the median temperature during the 10ps interval as a function of e-h density.
\begin{figure}
  \includegraphics[width=6cm]{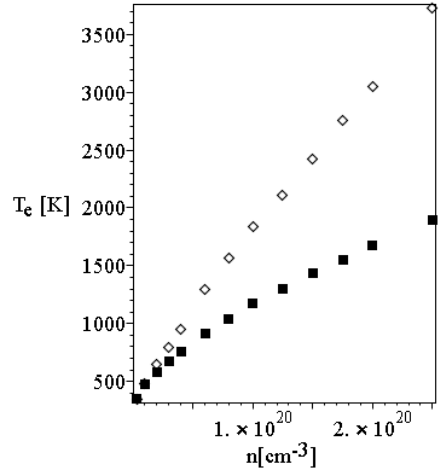}\\
  \caption{Mean carrier temperature in NaI within 10ps interval after photon absorption as a function of e-h track density: diamonds - $\Lambda=5$eV, solid boxes - $\Lambda=10$eV } \label{7}
 \end{figure}
 As seen from this figure the median carrier temperature during a 10ps non-proportionality stage is much higher for $\Lambda=5$eV in comparison with $\Lambda=10$eV because of less efficient energy out-flow.

 \quad Fig.8 illustrates the dependence of the fraction of carriers surviving non-linear quenching as a function of density, calculated with the use of (\ref{mobility}) and (\ref{QF}), where we substituted values of carrier temperature for each density.
\begin{figure}
  \includegraphics[width=7cm]{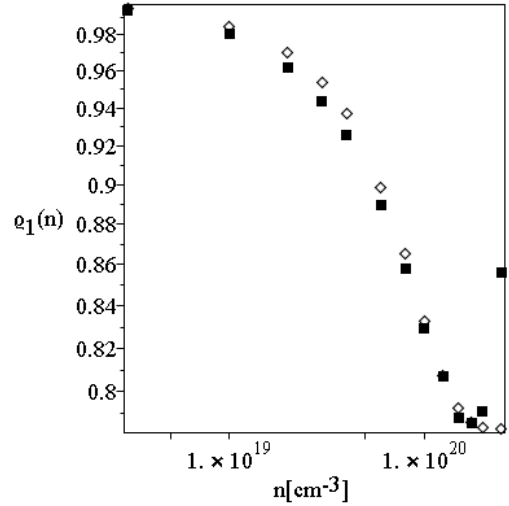}\\
  \caption{$\varrho_1(n)$ as a function of e-h track density: diamonds - $\Lambda=5$eV, solid boxes - $\Lambda=10$eV  }\label{8}
\end{figure}
The results shown in Fig.8 were obtained by calculating the ambipolar diffusion from carrier mobilities using the Einstein relation. The mobilities were calculated accounting only for phonon scattering which is the dominant contribution at high temperatures. Any contribution from impurity scattering was neglected. With the increase in carrier density the fraction of surviving carriers $\varrho_1(n)$ decreases. With no other fitting parameters, except values of deformation potential, we obtain the curves shown in Fig.8 corresponding to the middle of the range calculated by Li $\emph{et al}$ for $\mu_e=\mu_h$ within the range from 0 to 800 cm$^2$/V$\cdot$s\cite{Li}. The closest to both curves shown in Fig.8  is the curve from Li $\emph{et al}$ at $\mu$=2cm$^2$/V$\cdot$s. However, the latter curve was calculated with the mobilities and diffusion coefficients of electrons and holes being constant throughout the whole range of density variation. For a non-thermalised e-h gas, all transport parameters change independently. Fig.9 illustrates the behaviour of the normalised ambipolar diffusion coefficient. It is seen that the  ambipolar diffusion coefficient varies significantly. In addition, this variation is non-monotonic showing an increase in diffusivity  in the high temperature region.
\begin{figure}
  \includegraphics[width=8cm]{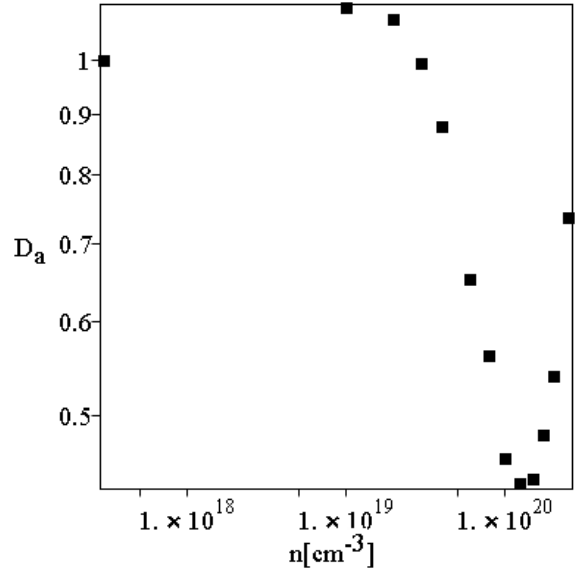}\\
  \caption{Normalised ambipolar coefficient of non-thermalised carriers as a function of carrier density in the track, $\Lambda=5$eV}\label{9}
\end{figure}
A non-monotonic variation of the ambipolar diffusion coefficient occurs in the high density and temperature region and indeed these  two factors may be of major importance for this region. High density results in stronger screening and increased carrier diffusivity. At high temperatures, the thermal momentum
of carriers exceeds the maximum phonon momentum setting the upper limits of all integrals over the phonon distribution corresponding to the Brillouin zone boundary. This introduces some extra temperature dependencies into the expression for mobility (\ref{mobility}). An increase in the  high temperature ambipolar diffusion coefficients results in either saturation or an increase of $\varrho_1(n)$ over the high density region shown in Fig.8. This feature is not present in any other models\cite{Bizarri,Li,payne1,payne2} of non-proportional light yield and may therefore  serve as an indicator of the non-equilibrated distribution of electrons, holes and phonons.

\quad Knowing the carrier temperature as a function of e-h track density allows us to evaluate the second factor in the expression for the activator capture efficiency, $\varrho_2(n)$, describing the fraction of carriers surviving exciton formation while migrating towards an activator. The Onsager radius is a function of carrier temperature, which in general strongly differs from the bath temperature, as seen in Fig.7. For the average initial electron-hole separation in the region of the track we choose $r_{e/h}=(2n)^{-1/3}$. This is an assumption which is consistent with the currently used model of a cylindrical track of varying density. This expression for $r_{e/h}$ differs from $\displaystyle{r_{e/h}\simeq3£E_g/\frac{dE}{dx}}$ used by Payne $\emph{et al}$\cite{payne1,payne2}, which describes the mean distance that  a  photoelectron passes to deposit sufficient energy to generate  a single e-h pair rather than e-h separation. Fig.10 shows the results of our calculations.
\begin{figure}
  \includegraphics[width=8cm]{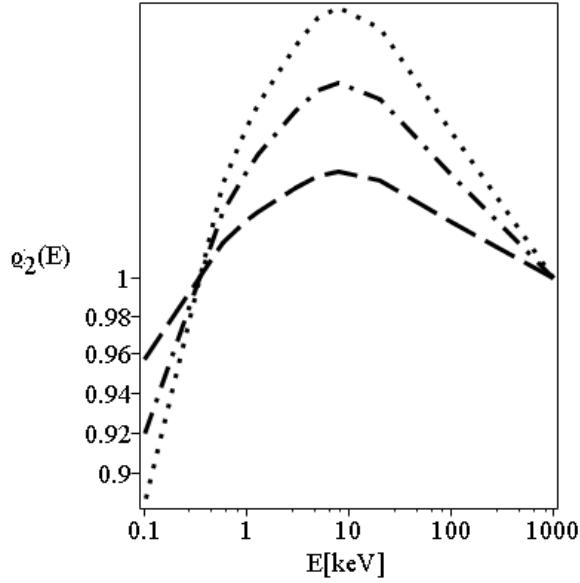}\\
  \caption{$\varrho_2(E)$ as a function of photoelectron energy, $\Lambda=10$eV : dash - $\eta_{e/h}=0.5$, dash-dot - $\eta_{e/h}=0.8$, dot - $\eta_{e/h}=1$}\label{10}
\end{figure}
To plot  this figure for $\varrho_2(E)$ as a function of photo-electron energy $E$  instead of $\varrho_2(n)$ we used the conversion between $E$ and the local e-h density in NaI derived by Vasiliev\cite{Bizarri}.

\quad The shape of  $\varrho_2(E)$ shown in Fig.10 is solely connected with the non-equilibrium distribution of carriers through the dependence of the Onsager radius on carrier temperature. It is different in any other models of non-proportionality in scintillator response. In particular in the papers by Payne $\emph{et al}$\cite{payne1,payne2}, $r_{Ons}$ is constant at $T=300K$, while $r_{e/h}$ decreases\cite{Bizarri} over the whole range 0.1-1000keV of $E$ thus making the function $\varrho_2(E)$ monotonic. The monotonic variation of the indepedent nonradiative fraction, "INF", was also reported by Li $\emph{et al}$\cite{Li}. In their expression for the simulated local light yield the factor $1-INF$ plays a role which is similar to $\varrho_2(E)$. Thus, both factors  $\varrho_1(E)$ and $\varrho_2(E)$ have features which are specific only for the non-equilibrium scenario.

\quad Finally, combining all our results together we obtain $\varrho(E)$ as shown in Fig.11.
\begin{figure}
  \includegraphics[width=12cm]{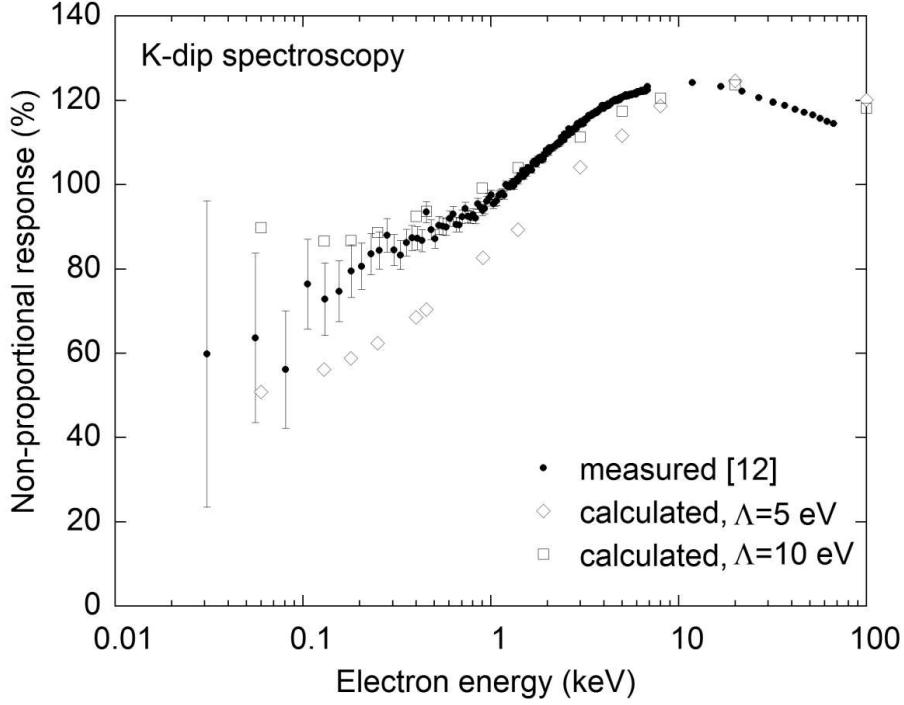}\\
  \caption{$\varrho(E)$ as a function of photoelectron energy, $\eta_{e/h}=1$: solid boxes - $\Lambda=10$eV, diamonds - $\Lambda=5$eV.
  Simulated curves are  superimposed on experimental data of I.Khodyuk $\emph{et al}$\cite{Khodyuk1} (adapted from Fig.11). Reprinted with permission from [I.V. Khodyuk, P.A. Rodnyi, and P. Dorenbos, J. Appl. Phys. $\textbf{107}$, 113513 (2010)]. Copyright [2010], American Institute of Physics.}\label{11}
\end{figure}

\quad The "hump" structure  is present in all simulations with different values of the parameter $\eta_{e/h}$. We see that as the exciton yield, $1-\eta_{e/h}$, increases the maximum value of $\varrho(E)$ decreases. However, the shape of the curves remain the same. The saturation of the non-proportionality factor $\varrho(E)$ at lower energies, or even the slight rise as seen for the case $\Lambda=10$eV, is evident in both simulated curves in Fig.11 and experimental data below 10keV. In simulated curves this is a specific feature of the non-equilibrium carriers arising from the special dependence of the ambipolar coefficient at high densities (temperatures) and corresponding features in $\varrho(E)$ at lower photo-electron energies. In fact particularly good agreement  both for the shape and the amplitude of the curve on the left from its maximum is achieved in the range of deformation potential of 7-8 eV. The simulated curves deviate from experiment on the right of the maximum. We emphasize that we did not attempt the best fit, but simply took the important parameters $a$, $\tau$ and $\bar{k}_2$ as they are quoted in the paper by Li $\emph{et al}$\cite{Li}. For example, increasing $a$ means that density of electronic excitations for any energy of photoelectron is smaller than what has been used in our modelling. This alone will result in a shift of the 
simulated curves in Fig.11 towards the  left, because the same density will now correspond to a lower photoelectron energy.
 It is likely that the parameters $\eta_{e/h}$ and $\bar{k}_2$ are functions of the non-equilibrium carrier temperature. In particular $\eta_{e/h}(E)<1$ for higher $E$ and carrier temperatures approaching room temperature. With density and temperature increasing, while $E$ decreases, we expect $\eta_{e/h}(E)\rightarrow£1$ for $T\gg£300$K. Therefore, accounting for this dependence will result in an increase in the magnitude of $\varrho_{max}$ as well as a corresponding increase in $\varrho(E)$ in the lower $E$ part of the curves shown in Fig.11. Overall, however, it will not affect the trend for saturation.

\quad Finally, the saturation or even rise of $\varrho(E)$ at lower energies is preceded by a change of curvature of $\varrho(E)$ on the lower energy side of its maximum. This change is not predicted by other models, although a change of curvature at lower energies was experimentally observed in NaI by Khodyuk $\emph{et al}$\cite{Khodyuk1}. Our modelling is consistent with the shape of their curve and the observed magnitude of non-proportionality, indicating that potentially the proof of non-equilibrium carrier effects was obtained in this experiment.
\section{Discussion}
\subsection{Thermalised model}
\quad  To discuss the implications of a non-equilibrium model, we produce formal calculations of non-proportionality for the fast (
much shorter than 10ps) thermalisation of carriers.  We first check the degree of degeneracy of carriers thermalised down to room temperatures at high electron and hole densities. Even at $n=1\cdot10^{20}$cm$^{-3}$, the electrons are strongly degenerate at $T_0=300$K in all crystals. For all of them we have $\zeta_e/T_0\geq£10$, where $\zeta_e$ is the chemical potential of electrons. The Einstein relation for an arbitrary degree of degeneracy\cite{Landau&Lifshitz} is $\displaystyle{eD_{e(h)}=\mu^*_{e(h)}n\frac{d\zeta_{e(h)}}{dn}}$. Correspondingly, for strongly degenerate carriers the limiting value of the factor $\displaystyle{n\frac{d\zeta_{e(h)}}{dn}}$ is $2/3\zeta_{e(h)}$, and the electron mobility will be closer to the value that we previously estimated for temperatures exceeding 1000K, under the assumption of non-degeneracy. In any case the mobility of thermalised electrons remains much higher than that of thermalised holes with a moderate degree of degeneracy.  As a consequence,  the electrons have a small effect on the ambipolar diffusion coefficient, which will therefore be determined  mostly by the holes. Heavy holes in NaI and CsI will have $\zeta_h\simeq£0$, so that
the factor $\displaystyle{n\frac{d\zeta_{h}}{dn}}\approx1.4T_0$ making ambipolar diffusion faster than our estimate, based on the assumption of hole non-degeneracy. Therefore, we will obtain the lower value of the non-proportionality of response if we use our results in expression (\ref{QF}) taking $T_e=300$K.

\quad Fig.12 shows the comparison of the factors $\varrho_1(n)$ for both thermalised and non-thermalised distributions. The upper curve was calculated for $\Lambda$-5eV. The curve for $\Lambda$=10eV essentially coinsides with it, and is not shown for clarity. The important point to note is that both curves show  signs of saturation at high densities. However, the increase in $\varrho_1(n)$ for the high $n$ end seen in Fig.8 for the non-thermalised distribution at $\Lambda$=10eV is absent at room temperature. This indicates that the increase in diffusivity at high track densities for $\Lambda$=10eV is due to extra temperature dependencies in the mobility of carriers arising from limitation of the integration over phonon momenta by Brillouin zone boundary rather than twice the thermal momentum of carriers. As seen by comparing the two curves in Figs.12, if the thermalisation is fast and carriers thermalise before nonlinear quenching sets in, then the quenching factor $QF=1-\varrho_1(n)$  becomes small. This simply reflects the result of increasing the mobility of carriers when the temperature decreases.
\begin{figure}
  \includegraphics[width=7cm]{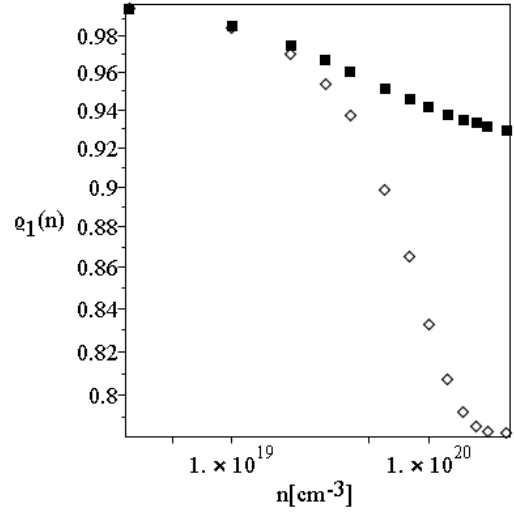}\\
  \caption{$\varrho_1(n)$ as a function of track density, $\Lambda=5$eV: solid boxes - thermalised distribution, diamonds - non-equilibrium distribution}\label{12}
\end{figure}
The second factor $\varrho_2(E)$ for this model is simply $\varrho_2(E)=1-\eta_{e/h}\exp\left(-r_{Ons}(T_0)/r({e/h}(E)\right)$. With $r_{Ons}$ being fixed at room temperature $\varrho_2(E)$ monotonically decreases with $E$ as in other models\cite{payne1,payne2,Li} within approximately the same range. To obtain a significant decrease in simulated light yield in this case the non-linear quenching must be pretty strong. This is not the case in our calculations. Fig.13 shows the calculated non-proportionality factors within our model for both the thermalised and non-thermal cases.
\begin{figure}
  \includegraphics[width=7cm]{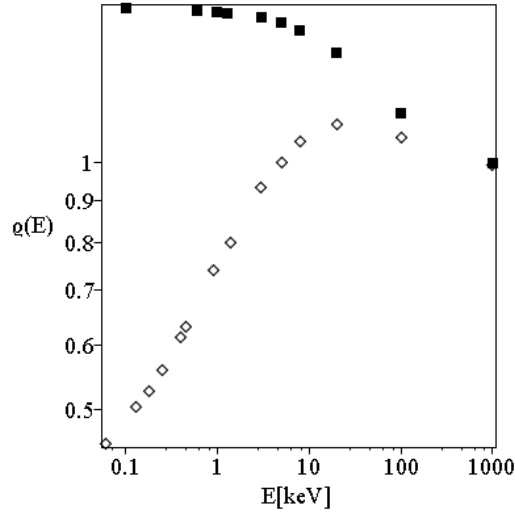}\\
  \caption{$\varrho(E)$ as a function of photoelectron energy, $\eta_{e/h}=1$, $\Lambda=5$eV: solid boxes - thermalised distribution, diamonds - non-thermalised distribution}\label{13}
\end{figure}
Overall the model of non-thermalised transport looks more realistic.

\quad The use of more definitive expressions for mobilities and diffusion coefficients, and the
Einstein relation in general, surprisingly becomes important in view of recent work on ambipolar
diffusion in the electron-hole plasma\cite{Efros}. It turns out that the ambipolar diffusion coefficient in a quasi-neutral e-h plasma differs from that
derived from the Einstein relation in the situation when the electron gas cannot
be viewed as ideal. The parameter which determines the role of the interaction between carriers in a charged degenerate plasma
is $r_s = (3/4\pi£na^3)^{1/3}$, where $a_B = \hbar^2\varepsilon_0/m^*_{e(h)}e^2$ is the Bohr radius. With the densities of carriers in the photoelectron cylindrical track in the range close to $10^{20}$cm$^{-3}$ we obtain
$r_s\geq 1,$ and therefore the thermalised degenerate gas of carriers inside the track is non-ideal. For example, in CsI with
$m^*_e= 0.235m$ and $\varepsilon_0= 5.8$ for $n = 1.0\cdot10^{20}$cm$^{-3}$ we obtain $r_s\simeq1$. Therefore, in a thermalised degenerate e-h plasma inside the track, the strong exchange and correlation results in a completely different ambipolar diffusion\cite{Efros}. At much higher transient temperatures of a non-thermalised e-h plasma, the kinetic energy of non-degenerate electrons and holes  exceeds the mean Coulomb energy and the e-h plasma will be close to ideal, validating the use of the standard Einstein relation.

\subsection{Temperature dependence of non-proportionality}
\quad In recent experiments, Khodyuk $\emph{et al}$\cite{Ivan} demonstrated a strong temperature dependence of non-proportionality in LaBr$_3$:$5\%$Ce for  thermal bath temperatures ranging from 450 to 78K. Below we discuss this result using both thermalised and non-equilibrium models.

\quad At first sight the rapid thermalisation of carriers leading to an increase in their mobilities with decreasing temperature seems to be consistent with experiment. Thermalised carriers diffuse away from the excited volume faster, and non-linear quenching is less effective. However, thermalisation down to room temperature over a 10ps time interval is already a problem unless crystals exhibit anomalously high anharmonic interactions. The problem is worse for thermalisation down to 78K. Below $T\simeq\hbar\Omega_{LO}$ all dissipation rates become strong functions of temperature, and temperature relaxation slows down significantly. Another argument is that below room temperature the mobility becomes dependent on the level of impurities, saturating when the temperature decreases. With typical activator densities in extrinsic scintillators it would not be surprising to have mobilities saturating  long before liquid nitrogen temperatures. Experimental data nonetheless demonstrates a strong variation in the non-proportional response in LaBr$_3$:Ce from 300 to 78K(see Fig.1 of ref\cite{Ivan}).

\quad A non-equilibrium model must also connect the observed dependence of the non-proportional response on ambient temperature with carrier mobility. However this connection is not direct, but comes through the potential dependence of varying temperature profiles in Figs.3-6 on ambient temperature. The dependence on ambient temperature affects the relaxation profiles through the details of energy dissipation from LO- and LA- phonons in the excited volume. If the thermalisation rate is small, as in Figs.3(c)-(d), 4(c)-(d), 5(c) and 6(a) and (c), the profiles of relaxing temperature are not sensitive to the variation with ambient temperature. This is expected, because over a 10ps time interval the temperature of carriers remains well above $T_0$, so that changing $T_0$ has a minor influence. However, in the situation of a more effective anharmonic decay, or higher group velocities for optical phonons allowing them to escape the excited volume, decreasing the ambient temperature means decreasing the relaxing temperature along the whole 10ps interval. To illustrate how this may affect  non-proportionality  in Fig.14 we show a comparison of the profile of relaxing the temperature for NaI with the profile calculated with the same decay and power dissipation parameters with the only the ambient temperature  $T_0$ decreased to 78K.
\begin{figure}
  \includegraphics[width=6cm]{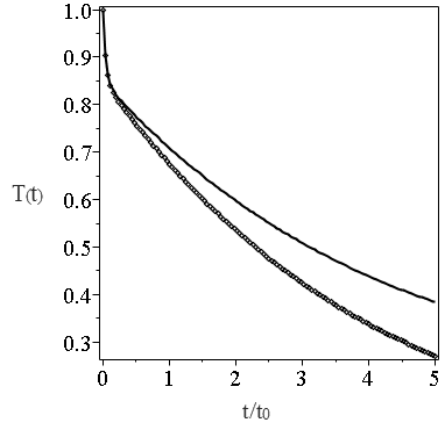}\\
  \caption{Temperature evolution in NaI: $n=1\cdot10^{20}$cm$^{-3}$, $\gamma_{LA}t_0=1$, $\Lambda=10$eV. Upper curve - $T_0=300$K, lower curve - $T_0=78$K }\label{14}
\end{figure}

As is seen in Fig.14, the difference between $T_e(t)$ profiles for $T_0=300$K and $T_0=78$K reaches 200K by the end of a 5ps interval. Correspondingly, the effective diffusion coefficient for the lower curve is larger than for the upper curve. It is worth noting that the decay time $\gamma_{LA}^{-1}$ was chosen to be equal to $t_0=2.1$ps. Ballistic LA-phonons in NaI pass a distance of 7nm on average over $t_0$. Thus, the effective "decay" rate of LA-phonons in the balance equations, $\gamma_{LA}t_0\cong£1$, describing the decrease of numbers of LA-phonons in the excited volume may be consistent with the flux of outgoing LA-phonons. The calculation  of non-proportionality for the lower ambient temperature of $T_0=78$K  taking the effective temperature of carriers to be 15$\%$ lower than that of the scintillator at room temperature yields $\varrho(0.5\mathrm{keV})$=0.98 as opposed to 0.89 for $T_0=300$K - quite a significant variation. Clearly, the effect is dependent on material parameters describing energy exchange between the excited volume and its environment.

\quad The effect of bath temperature on the non-proportional response in doped alkali halide scintillators is of special interest to study experimentally. The exponential dependence on effective temperature in $\varrho_2$ may significantly enhance small variations in the non-equilibrium temperature of the carriers due to crystal cooling in comparison with scintillators, which do not exhibit the "hump" in non-proportional response, such as LaBr$_3$:Ce observed by Khodyuk $\emph{et al}$\cite{Ivan}. Under such conditions non-thermalised transport can cause an anomalously strong temperature dependence of non-proportional response in alkali halide scintillators. The experimental observation of such an anomaly could be a strong argument in favor of the non-equilibrium scenario.

\subsection{The effect of the non-equilibrium state on quenching characteristics}

\quad So far we have discussed the implications of the carriers being non-thermalised during the non-proportionality stage by comparing carrier transport characteristics. Apart from transport properties,  the effectiveness of non-linear interactions resulting in quenching of luminescence is determined by the dependence of various interaction constants on the mean energy of carriers and also on such parameters as $\eta_{e/h}$ (discussed in section IV) determining the exciton yield. Therefore the concept of non-thermalised carriers in relation to  non-proportionality in scintillators is more general than the concept of non-thermalised carrier transport. In particular, non-proportional response will depend both on the carrier diffusion coefficients and the non-linear interaction constants changing with time.

\quad It is well known that in equilibrium conditions the Auger recombination rates may show a significant dependence on temperature\cite{Landsberg}, supporting our arguments. Similarly, the bimolecular quenching parameter $k_2(t)$ may also depend on varying carrier energy, which   adds an extra transient variability to luminescence quenching. If such a dependence is present one should be able to experimentally distinguish between the scenario of distributions of non-thermalised carriers  and that of carriers which relaxed their kinetic energy down to the temperature of the thermal bath, for both the same excessive non-equilibrated number.

\subsection{K-dip spectroscopy and non-thermalised electrons and holes}

\quad The role of non-thermalised carriers in scintillator non-proportionality is most important in the parts of the photoelectron track with the
highest density. In ordinary situations this is the part of the cylindrical photo-electron track closest to the end where ionisation is the most intensive. One promising new experimental technique for the identification of the non-thermalised carrier scenario, is K-dip spectroscopy\cite{Dorenbos2}. In K-dip spectroscopy, the energy of incident X-rays, $E_X$, is chosen to be close to the ionisation energy of the  K-shell, $E_K$. As a result the K-photoelectron is released with an energy of $E_X-E_K$, while the K-shell hole left behind may result either in emission of escape X-rays due to $K_{\alpha,\beta}$, $L_{\alpha,\beta}$, $M_{\alpha,\beta}$ transitions or Auger recombination, leaving holes in outer shells and generating the Auger electrons. Neglecting the escape peaks, we concentrate on energy deposition in the K-cascade, which is the process of sequential Auger recombinations of the holes in K-, L- , M-... shells until holes start to be created in the valence band. In the K-cascade the first Auger electrons originating from  $K_{\alpha,\beta}$, $L_{\alpha,\beta}$ and also $M_{\alpha,\beta}$ transitions possess an energy $\geq$ 1 keV. They all are released at the absorption site but move in random directions giving rise to their own cylindrical ionisation tracks. Due to the randomness in the directions of the Auger electrons ejection, these tracks do not overlap. The highest density of the generated electron-hole plasma will be located at the end points of each track. The Auger electrons with energies exceeding 1 keV have low ionisation cross-section, so that the starting points of their cylindrical tracks are well separated from the absorption point. However, in the last stages of the K-cascade, the Auger recombination involves the outer N- and O-shells, and the Auger electrons are ejected with energies less than 1keV. These electrons will create a high density electron-hole plasma in the vicinity of absorption site. If the energy of the K-photoelectron is less than 1 keV, its e-h ionisation cloud will be also located in the immediate vicinity of the absorption site.
For $E_X\gg£E_K$ the fraction of energy lost due to non-linear quenching in the  vicinity of the absorption site is small and can be neglected. However in the K-dip spectroscopy experiment with $E_X-E_K\rightarrow£0$ the highest e-h density region is created at the absorption site with the K-photoelectron ionisation cloud overlapping with those from the last few Auger electrons from the concluding stage of the K-cascade. t As a consequence, the density in this region may exceed the density of the e-h plasma at the ends of few cylindrical tracks. In this case the energy lost due to non-linear luminescence quenching in the central region may become comparable to, or even exceed, that  lost at the ends of the cylindrical tracks.

\quad The experimental observation of an "end-point", $E_X-E_K\rightarrow£0$, of non-proportionality in the K-dip spectroscopy experiment\cite{Dorenbos2} shows an increase in the  dispersion of  data points starting at $E_X-E_K\leq$1keV. Khodyuk $\emph{et al}$ attribute this  to the low accuracy of the measurement in this region, resulting primarily from the subtraction of two large, similar numbers and to systematic errors. We believe that this data spread is due to more fundamental reasons. Indeed, the degree of overlap of ionisation clouds in the vicinity of the excited atom (i.e., the absorption site) is subject to the randomness in the  velocity distributions  of the last few Auger electrons and the K-photoelectron. As a result, the actual density of e-h plasma in the vicinity of the absorption site will strongly fluctuate about its mean value giving rise to an additional  spread in the data points - smaller non-proportionality for lesser overlap, and stronger for higher overlap. Subtracting the mean values therefore reveals the data spread due to these fluctuations.

\quad From an experimental standpoint, we strongly recommend that this spread is studied in greater detail. With the mean e-h density in the vicinity of absorption site substantially exceeding that at the end points of the cylindrical tracks, the expected peak temperature $T(0)$ may show a significant dependence on the e-h density. Thus, the shape of the end point ($E_x-E_K$) of the non-proportionality curve in K-dip spectroscopy can serve to test for and potentially probe the effects of non-equilibrium carriers. Another possibility is to study the possible variation of this shape with ambient temperature, which will cause a change in the dynamics of hot carrier relaxation.
\section{Summary and conclusions}

\quad We have developed a non-equilibrium model for the  non-proportional response observed in scintillators. We complement previous models, based on the idea of fast carrier thermalisation and thermalised carrier transport, with a detailed analysis of energy exchange between carriers and phonons. The main conclusion drawn from this analysis is that unless the scintillator material possesses anomalously high anharmonicity, the energy flow from the excited volume containing e-h pairs generated by the absorbed X- or $\gamma$-ray photon, is impeded by the relatively immobile distribution of longitudinal optical phonons. Once the thermalisation processes have been studied in detail the best strategy for optimising scintillator response can be deduced. The present work indicates that the most promising direction to pursue will be to affect the thermalisation rate in order to create the conditions for faster out-diffusion of carriers from the initial high density ionization track. In practice, this may require either making thermalization faster or slower, depending on temperature dependence of carrier mobility and coefficients of Auger recombination or bimolecular dipole-dipole F$\mathrm{\ddot{o}}$rster transfer mechanism of non-linear quenching. The rate of thermalisation may also be affected by creating a specific type of disorder and in this regard we note that scintillators need not be as structurally and stoichiometrically perfect as semiconductors to achieve good performance.

\quad We wish to thank I.Khodyuk for providing the data used in Fig.11.

 \end{document}